\documentclass[journal=jacsat,manuscript=article]{achemso}

\usepackage{hyperref}
\usepackage{amsmath}
\usepackage{bm}

\usepackage{tikz,xcolor,hyperref}

\definecolor{lime}{HTML}{A6CE39}
\DeclareRobustCommand{\orcidicon}{
\begin{tikzpicture}
\draw[lime, fill=lime] (0,0)
circle[radius=0.16]
node[white]{{\fontfamily{qag}\selectfont \tiny \.{I}D}}; 

\end{tikzpicture}

\hspace{-2mm}
}
\foreach \x in {A, ..., Z}{%
\expandafter\xdef\csname orcid\x\endcsname{\noexpand\href{https://orcid.org/\csname orcidauthor\x\endcsname}{\noexpand\orcidicon}}
}

\usepackage{chemformula} 
\usepackage[T1]{fontenc} 

\author{Changlin Wu}

\affiliation[Unknown University]
{Key Laboratory of Terahertz Solid-State Technology, Shanghai Institute of Microsystem and Information Technology, Chinese Academy of Sciences, 865 Changning Road, Shanghai 200050, China}
\alsoaffiliation[Second University]
{Center of Materials Science and Optoelectronics Engineering, University of Chinese Academy of Sciences, Beijing 100049, China}

\author{Chang Wang}

\affiliation[Unknown University]
{Key Laboratory of Terahertz Solid-State Technology, Shanghai Institute of Microsystem and Information Technology, Chinese Academy of Sciences, 865 Changning Road, Shanghai 200050, China}
\alsoaffiliation[Second University]
{Center of Materials Science and Optoelectronics Engineering, University of Chinese Academy of Sciences, Beijing 100049, China}
\email{cwang@mail.sim.ac.cn}

\author{Guanjun You}

\affiliation[Unknown University]
{Shanghai Key Lab of Modern Optical Systems, Terahertz Technology Innovation Research Institute, and Engineering Research Center of Optical Instrument and System, Ministry of Education, University of Shanghai for Science and Technology, Shanghai 200093, China}
\email{gjyou@usst.edu.cn}

\author{Juncheng Cao}

\affiliation[Unknown University]
{Key Laboratory of Terahertz Solid-State Technology, Shanghai Institute of Microsystem and Information Technology, Chinese Academy of Sciences, 865 Changning Road, Shanghai 200050, China}
\alsoaffiliation[Second University]
{Center of Materials Science and Optoelectronics Engineering, University of Chinese Academy of Sciences, Beijing 100049, China}

\title[An \textsf{achemso} MS-SNOM]
  {THz Nanoscopy of Metal and Gallium Implanted Silicon}

\abbreviations{THz, TDS, NF, s-SNOM, SPM}

\begin{document}


\begin{figure}
  \includegraphics{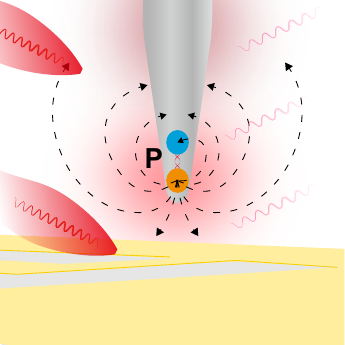}
  \caption*{Tip apex source dipole radiating over samples.}
  \label{TOC}
\end{figure}


\begin{abstract}
  Drude model successfully quantifies the optical constants for bulk matter, but it is not suitable for subwavelength objects. In this paper, terahertz near-field optical microscopy and finite element simulation are used to study gold patches fabricated by Gallium etching. Electron transport is discovered in determining the optical signal strength. The signal from substrate is more complicated and still not fully understood. As the etching area decreases, near-field interaction is not dominated by doping concentration, and a higher signal is observed near connected metals. With the help of simulation, the abnormal enhancement phenomenon is discussed in detail, which lays the foundation for further experimental verification. 
  
\end{abstract}


\section{Introduction}
  Quantifying the optical properties of terahertz (THz) subwavelength objects is an important and difficult task. In most frequency spectra, the optical properties are definite\cite{palik1998handbook}. THz time-domain spectroscopy (THz-TDS) based on photoconducting dipole antennas was invented and used in the measurement of bulk materials\cite{auston1988electrooptical, smith1988subpicosecond}. THz wave band is opaque to metals and water, strongly absorbed by some molecules\cite{gallot1999measurements, cheon2016terahertz} and transparent to most dielectric materials. So that THz imaging\cite{hu1995imaging, wang2022overview, qiu2022terahertz, maissen2019probes, mastel2018understanding} and fingerprint technology\cite{ walther2002collective, shen2021ultra} are appealing for wide usage in the future. THz waves are sensitive to low-energy phenomena in bulk matter and molecules. These quantum phenomena can be simplified into classic oscillator models, like Drude model\cite{ordal1983optical, benavides2016numerical} and Drude-Lorentz model\cite{jeon2004optical}. 
  
  Recent studies found that the optical constants in the THz range are different, in which a 3 orders discrepancy between experimental and Drude model results was reported\cite{pandey2016non}. The size and shape of matter are vital at THz frequencies as the free electron gas assumption is invalid, and we have to analyze the inner electrical interaction\cite{park2009terahertz, chen2013atomic}. A calculation method that adapts to the complex boundary conditions must be introduced\cite{benavides2016numerical}. 
 Additionally, subwavelength resolution optical equipment is needed to extract constants, such as scattering-type scanning near-field optical microscopy (s-SNOM) based on THz-TDS setup\cite{moon2015subsurface, chen2020thz, cocker2021nanoscale}. Atomic force microscopy (AFM) probe produces strong electric field enhancement at tip apex and radiates the near-field signal to the detector\cite{2008Terahertz, hillenbrand2002phonon, cvitkovic2007analytical, knoll2000enhanced}. Sub- $10~\rm{nm}$ imaging resolution have been achieved\cite{chen2019modern, pizzuto2021anomalous, zhang2021direct}. 
 
 However, the nonlinear effect makes the optical properties not proportional to detected signal. The signal from each pixel is correlated to the background, and noise is inevitably introduced in demodulation. It is a feasible alternative to measure and compare samples of known materials, qualitatively reflecting the relative relation of the dielectric constants. However, as we previously mentioned, the known materials may respond differently in the THz range, making the comparison results less reliable. 
 Combine finite element method (FEM) simulation with experimental data in THz-Nano matter interactions is essential\cite{huth2013resonant, chen2017rigorous, mcardle2020near, chen2021hybrid,mcleod2014model, mooshammer2020quantifying, feres2021sub}. 
 The FEM simulation result is based on Maxwell's equations, which ensures a fixed field strength ratio for different incident electric field strengths. Indeed, the experimental data of optical signals are also a relative value. To eliminate the difference between simulations and experiments, we need to modify simulation parameters until reaching the experimental results.
 
  Here, we carry out a fast FEM simulation method and transform all pixels to near-field images (NFIs), which reduce the influence of computational error fluctuations on the result.
 Micron and nanoscale samples, as connected patches and isolating patches on high-resistance Si-substrate, are fabricated by a focused ion beam (FIB) and measured via THz-TDS s-SNOM. 
 Compared with simulated images (SIMUs), the micron sample (1) results have good consistency, but the nanoscale sample (2) exhibits some unexplained experimental phenomena. We speculate that the abnormal enhancement on the substrate is related to both surface roughness and signal enhancement caused by THz light irradiation. A proper experiment is needed to explain these phenomena.
 
\section{Results and Discusses}

\begin{figure}
  \includegraphics{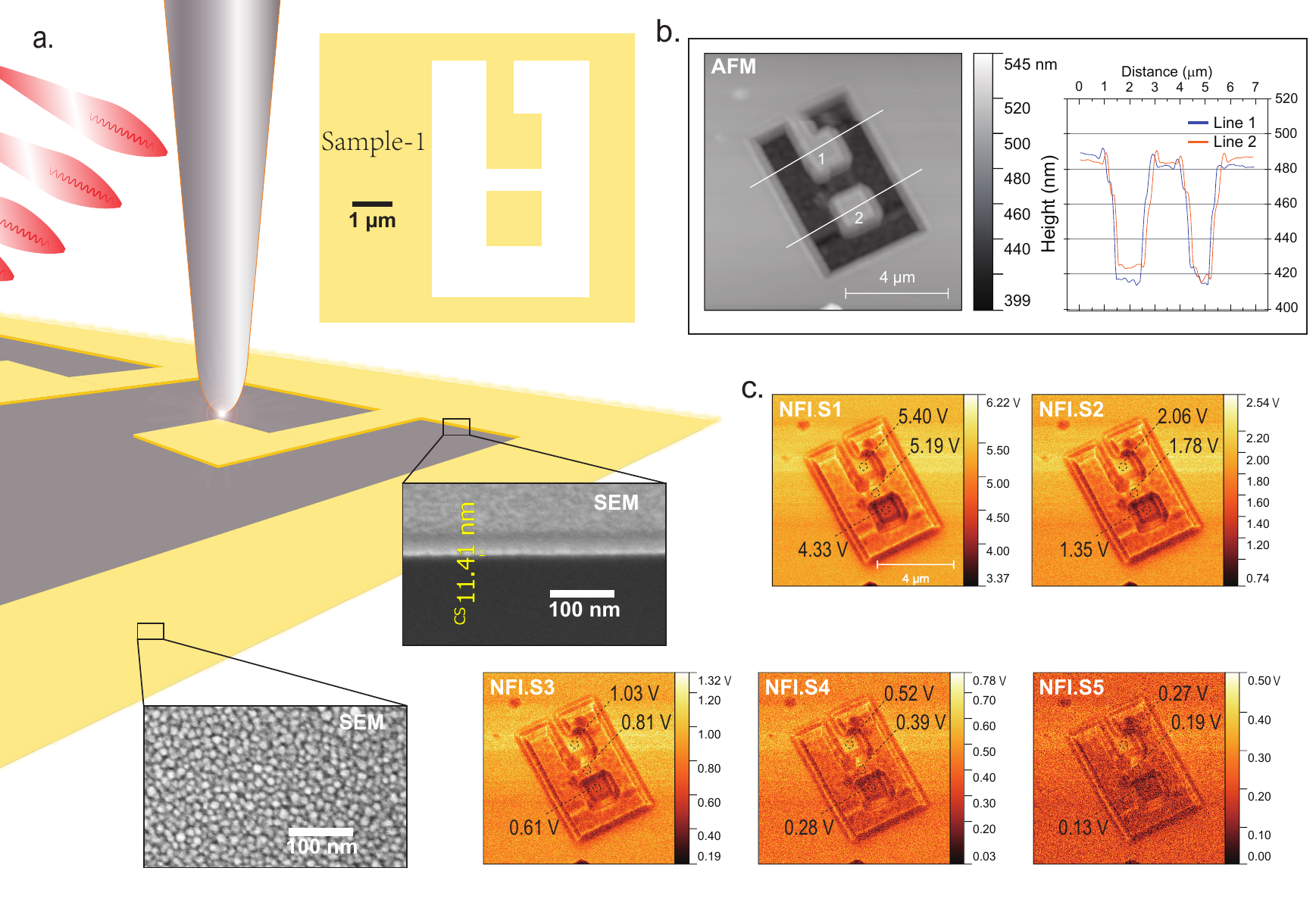}
  \caption{THz-TDS s-SNOM sample test. (a.) Experimental setup and Sample-1. (b.) AFM topography of the connected and isolated patches. (c.) Near-field imaging with the S1-S5 order optical signal.}
  \label{fig1}
\end{figure}

  THz-TDS s-SNOM is a combined tool for acquiring optical signals and topography. THz pulses are focused on the tip-sample junction with p-polarized electric field, and a dipole response is induced in the AFM probe. In a simplified consideration, the source dipole is locates at the tip apex, moves, irradiates the underlying sample structure and reflects high-order signals ($S_1, S_2, S_3, S_4, S_5$) to the far-field. The samples are gold films on high-resistance silicon with structures fabricated by state-of-the-art FIB techniques. As depicted in Fig. \ref{fig1}(a.), patches are dug up from the whole gold film and some patches will be connected. The different electrically connected states will lead to strength variation of the optical signal\cite{chen2020thz}. 
  On Sample-1, there are gentle channels down to tens of nanometers into the silicon substrate, forming two patches (see  Fig. \ref{fig1}(b.)). Optical signals $S_1$ to $S_5$ are collected in NFI.S1-S5. We compare the results of 3 points in the NFIs (averaged over $0.39~\rm{\mu m} \times 0.39~\rm{\mu m}$ area). The connected patch (upper point) has the highest signal strength, and the isolated patch (bottom point) has an extremely weak reflected signal even compared with the Si - substrate (middle point). NFI.S1-S2 have robust values but a larger background base level. NFI.S4-S5 have a nearly zero background base level but a significant noise disturbance. NFI.S3 is a suitable candidate for our discussion. As the background influence is equally applied to connected patches, isolated patches and Si - substrates. Assuming that the incident background electric field with frequency $\omega_0$ is $E_Z^{(0)} (\omega_0)$, the first-order signal $S_1=\alpha_1 E_Z^{(0)} (\omega_0\pm f_0)$ generated by AFM probe modulation with a coefficient of $\alpha_1$ and tapping frequency $f_0$. Second-order and higher-order signals result from the mixing of different frequencies as $S_n=\alpha_n E_Z^{(0)} (\omega_0\pm nf_0)$; here, $\alpha_n$ is the nonlinear modulation coefficient of different order signals. Although different orders of signals involve nonlinear optical processes, they are linearly correlated. In Sample-1, the signal strengths of each $S_n$ have the following relationship $2E(middle~point) = E(upper~point) + E(bottom~point)$. This is an unusual result because the signal from the Si substrate is very high. It is caused by the gallium ions implantation and resulted in electrical property changing of Si-substrate. Because we have etched down more than $40~\rm{nm}$, patches are electrically isolated to $\rm{Ga^{+}-Si}$ substrate\cite{zhang2021detection, graziosi2018single}. Without considering the substrate signal, we have the following relation in Sample-1
  \begin{equation}
  \begin{minipage}[c]{0.80\linewidth}
    \centering
    $\alpha_n^{(iso)}=\frac{1}{2} \alpha_n^{(con)}$
  \end{minipage}
  \label{eq1}
  \end{equation}
  
\begin{figure}
  \includegraphics{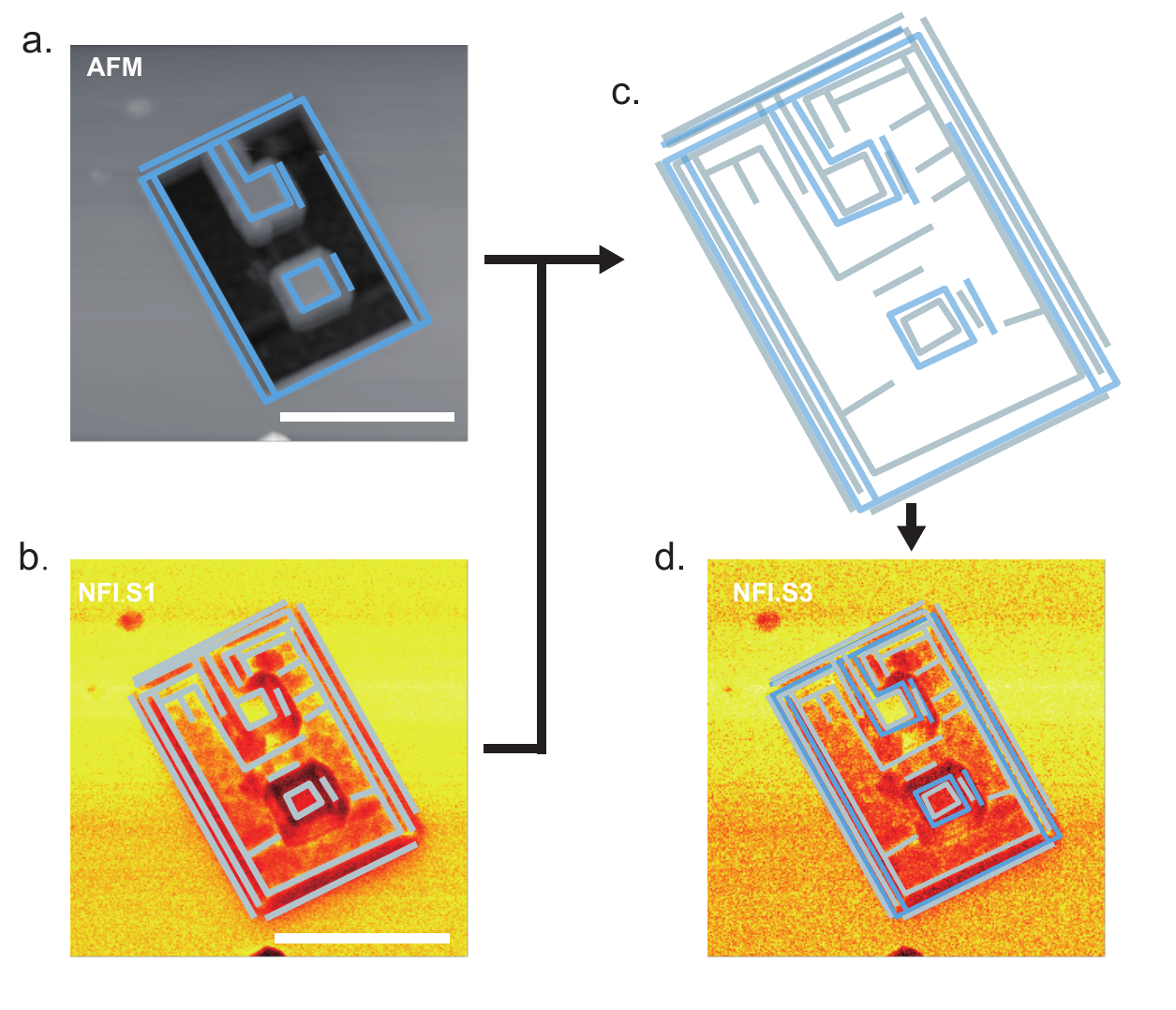}
  \caption{Extraction of the edges from AFM and SNOM images and comparison of them with NFI.S3. Scale bar $5~\rm{\mu m}$.}
  \label{fig2}
\end{figure}

  The sidewalls of Sample-1 are not steep ($\leq400~\rm{nm}$ in the lateral direction, $\leq100~\rm{nm}$ in the vertical direction), which is caused by the ion etching and results in material accumulation on the edges. Gallium mixed with silicon oxide and even gold will stack beside the channels, as we can see from the profiles along the whole samples (see Fig. \ref{fig1} (b.)). These hill-like structures weaken the near-field signal due to the small area for interaction with the source dipole at the tip apex. 
  However, the inclined slope will increase the contact area with the probe shaft, enlarge the plate area of the tip-sample capacitor, and radiate more high-order signals to the far field. As depicted in Fig. \ref{fig2}, topography ridges extracted from (a.) and boundaries of the high signal strength from (b.) are compared in (c.). The pattern of Fig. \ref{fig2}(c.) gives an overview of the relationships between geometrical boundaries to sharply changed near-field signal boundaries. Even though the pattern is extracted from NFI.S1, it highly fits with other NFIs results (see Fig. \ref{fig2}(d.)). The optical edges are mostly encircled by the geometrical edges but some wavy bottoms on the Si - substrate also have great optical signals. The ranges of the wavy structures are small ($\leq10~\rm{nm}$), which means that the topographical unevenness will cause a decisive signal difference. This effect may be stronger than the variation caused by the optical properties of the materials. The signal on the substrate near the connected patch is enhanced and we will discuss this further regarding Sample-2.
  
  The connected patch has higher signal strength than the isolated patch, which is solid evidence that the Drude model fails at subwavelength. The main reason for this phenomenon is the transport and balance of electrons, followed by the shape of conductors. FEM software COMSOL is introduced to investigate the electrical properties in simulation, and we lack proper experimental methods to verify them. 
  
\begin{figure}
  \includegraphics{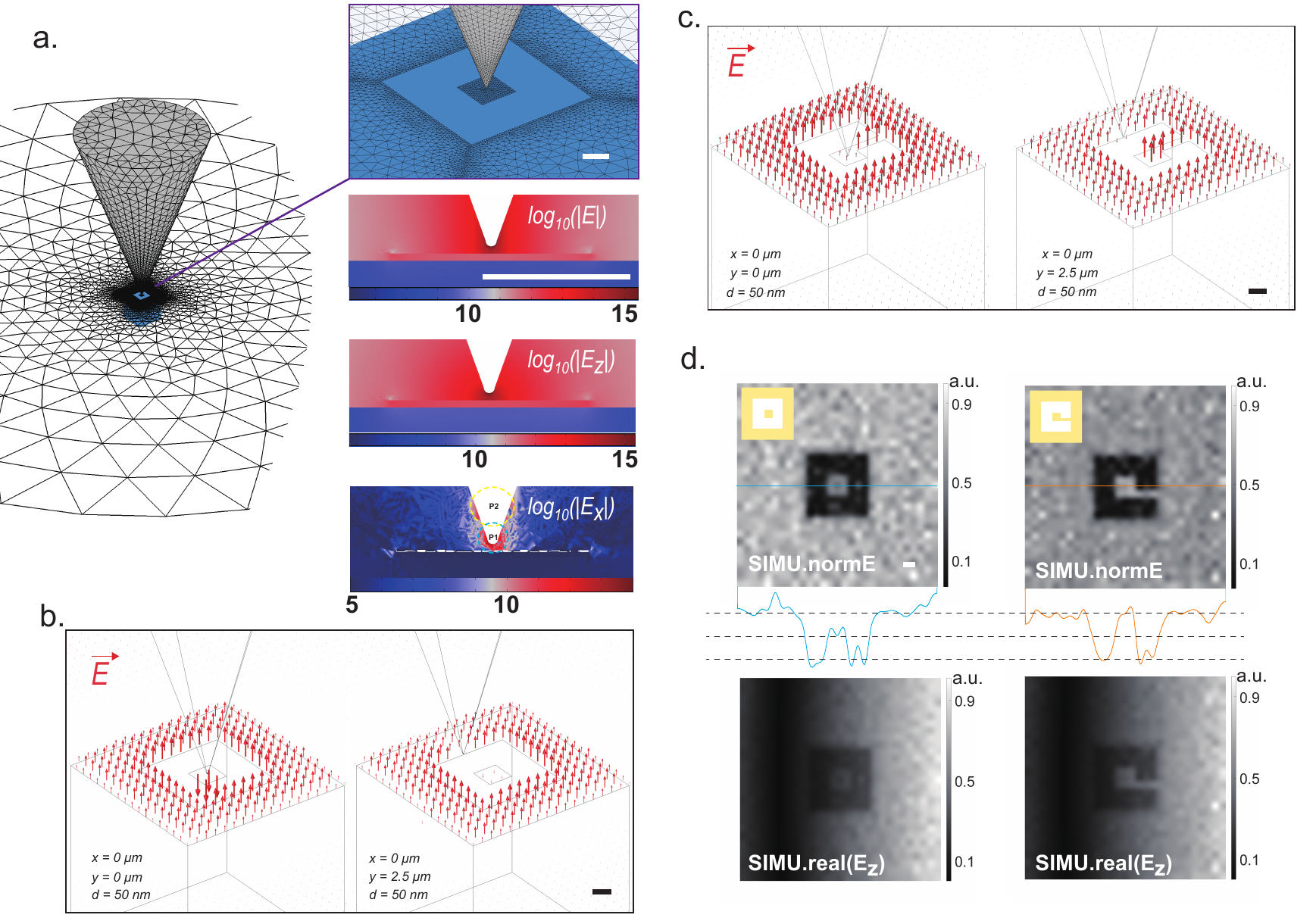}
  \caption{Finite elements simulation. (a) Models. (b) Electric field distribution of isolated patch. (c)  Electric field distribution of a connected patch. (d) Simulated images (grayscale normalized). Scale bar $1~\rm{\mu m}$.}
  \label{fig3}
\end{figure}

  Simulation plays a vital role in the study of s-SNOM. First, the physical model is transformed into meshing grids (see Fig.\ref{fig3} (a.)). The AFM tapping process is replaced with a changing tip-sample distance in the frequency domain. The electric field generation in the tip-sample junction is a nonlinear process, and the electric field is inversely proportional to the cubic power of the distance. However, photodetector will average the radiation signals among tapping amplitudes, resulting in a stable electrical signal when the amplitude is sufficiently large. In this experiment, electrical signals from THz-TDS s-SNOM are almost unchanged when the amplitude is larger than $100~\rm{nm}$. Therefore, a tip-sample distance of $50~\rm{nm}$ is used in the simulation to replace the Fourier transform in the radiation detection process. This method will reduce the amount of calculation by dozens of times, which is the key to realizing simulation of the NFIs. 
  
  An external electric field is applied in the Z-direction, which leads to the formation of source dipole $P1$ at the tip apex. This source dipole will excite charges on the probe shaft to produce dipole $P2$ with opposite polarization. In this way, additional dipoles are formed along the entire probe axis. One of the results is that the probe length greatly influences the near-field strength $|E_Z|$, while $|E_X|$ is relatively weak (see Fig. \ref{fig3}(a.)). Drawing the electric field vectors on the sample surface of the isolated patch (b.) and the connected patch (c.) at positions $(0~\rm{\mu m}, 0~\rm{\mu m})$ and $(0~\rm{\mu m}, 2.5~\rm{\mu m})$, all electric fields are perpendicular to the conductor surfaces of the probe and patches. The source dipole on the apex will have a strong electric field pointing down toward the underneath structure. The isolated patch is electrically independent of the peripheral conductor film and is dominantly influenced by the tip apex. The directions of the electric field vectors are different between the isolated and connected patches. This phenomenon is caused by the free charge density $\rho_f$ on the conductive surface of the specimen. According to the boundary conditions of ideal electrical conductor and assuming the background electric field points in the Z-direction, $\hat{n}\times (\vec{E}-0)=0$ and $\hat{n}\bullet (\varepsilon \vec{E}-0)=\rho_f$ hold on the surface.
  The perpendicular conductor is regarded as the ground with free charge density $\rho_f^{(c)}=\varepsilon_0 E_Z^{(0)}$. Here $\varepsilon_0$ is the vacuum permittivity. The isolated patch is polarized by the source dipole on tip apex, which results in negative free charge density $-\rho_f^{(i)}$. The sign is not important, because the frequency of THz waves is much higher than that of AFM tapping. All the detectable parameters are averaged over thousands of tapping cycles and related to the dependent variable $\rho_f$. Relative optical constants, such as the permittivity ratio $\varepsilon_{iso}/\varepsilon_{con}$, can be written in the expression 
  
  \begin{equation}
  \begin{minipage}[c]{0.80\linewidth}
    \centering
    $\varepsilon_{iso}/\varepsilon_{con}=\varepsilon_{patch}/\varepsilon_{Drude}=\frac{|\rho_f^{(i)}|}{ |\rho_f^{(c)}|}$
  \end{minipage}
  \label{eq2}
  \end{equation}
  
  Near-field results from experiments and simulations have shown that electron transport links electrical properties to optical properties. The free charge density excited by the source dipole will lead to a change in permittivity. Data of isolated patches and connected patches are obtained in the experiment, but their optical constants cannot be directly determined. In contrast, FEM simulation starts with modelling, and the electric field distribution results are obtained by setting optical constants. Comparing the experimental (see Fig. \ref{fig1}(c.)) and simulation (see Fig. \ref{fig3}(d.)) results can help us determine the optical constants. As depicted in Fig. \ref{fig3}(d.), the normE of isolated patches and that of connected patches are separately simulated (SIMU.normE, normalized). According to a comparisson of the data crossing centers of the profiles, the simulation highly agrees with experimental results of Eq. \ref{eq1}. 
  From another view, electrons are more likely to flow though connected patches and consume more THz energy. The absorption makes frequency red shift and results in stronger high-order signals $S_n$. Edge enhancement effects can be explained as more electrons accumulate on the edges and consume more energy than peripheral metal. However, because the simulated EM waves are stored in a complex data form, the light-and-shadow as shown in SIMU.real($\rm{E_z}$) is a problem that cannot be ignored.

\begin{figure}
  \includegraphics{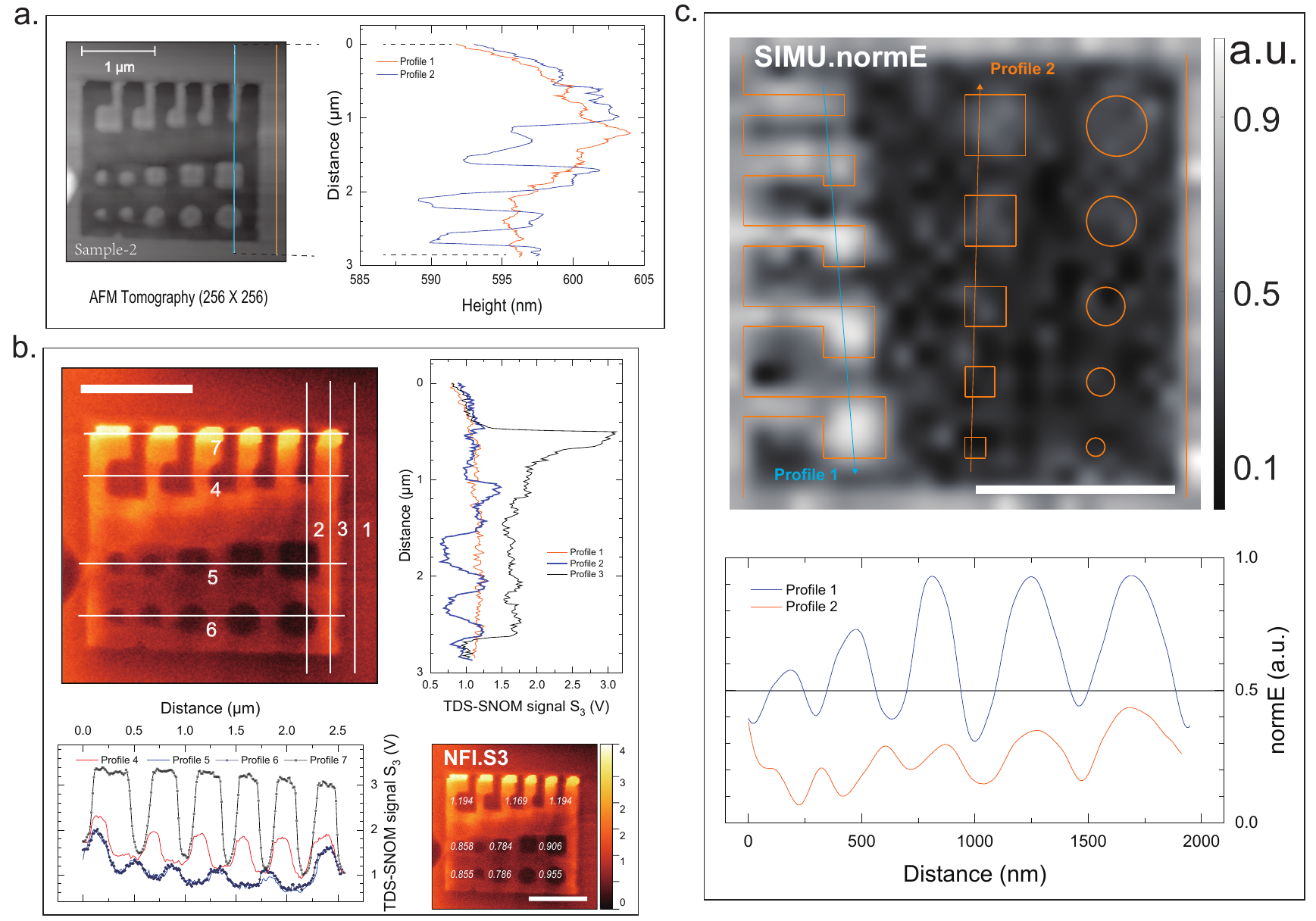}
  \caption{NFIs and SIMUs of Sample-2. (a.)AFM surface topography ($256\times256$ pixels, profiles 1 and 2). (b.) S3 near-field imaging ($256\times256$ pixels, profiles 1-7). (c.) Simulation results (step length $100~\rm{nm}$, profiles 1 and 2). Scale bar $1~\rm{\mu m}$.}
  \label{fig4}
\end{figure}

  The interaction between s-SNOM and nanoscale and microscale samples are different. Under the condition of extreme precision of the FIB, Sample-2 with a $100~\rm{nm}$ size and a $10~\rm{nm}$ etching depth is manufactured. The slight etching makes the metal directly connected with the Si-substrate. Sample-2 consists of three groups of patches ranging from $100~\rm{nm}$ to $300~\rm{nm}$ and shapes of circles or squares. A group of square patches are connected to the outside, and their connecting lines are all equal in length. There are small unevenness on the Si - substrate(see Fig. \ref{fig4}(a.)). As the FIB etching process focuses the ion beam on a square frame, the etching intensity will increase with the focused area decreases. Therefore, etching of the gold film will inevitably lead to different etching depths. We attribute the extremely high signal on the Si - substrate main to the increment of Gallium ion concentration. But, the doping is not the only reason caused the pattern signal difference. As depicted in Fig. \ref{fig4}(b.), signal strength on peripheral gold film is about $1.1~\rm{V}$, which is higher than that on the previous film (see Fig. \ref{fig1}). Profiles 1-7 are acquired from NFI.S3 for the gold film (profile 1), substrate (profiles 3 and 7) and patches (profiles 2, 4, 5, and 6). The upper substrate area has an extremely high signal strength reaching $3.0~\rm{V}$. The signal strength of the upper left area is stronger than the lower right area. This strange phenomenon is not caused by the unevenness of the sample surface or incident angle of the THz wave because profile 5 and profile 6 are highly coincident. Look at connected and isolated metal patches again for possible explanation. Under the same topography conditions, the signal around connected patches (Profile 4 in Fig. \ref{fig4}(b.)) is stronger than that around isolated patches (Profiles 5-6). This means patches are electrically connected to the doped substrate. As optical properties of doping substrate is unknown, we also use intrinsic Silicon instead here. 
  
  The irregular strength of different metal patches can be verified by simulation (see Fig. \ref{fig4}(c.), profile 1). A simple explanation is that the comparable patches’ sizes ($100~\rm{nm} ~-~300~\rm{nm}$) to tip apex diameter ($80~\rm{nm}$) will lead to free electron density change and antenna effects on the small metals. With the size shrinking, fewer free electrons are on the patches, while the electrons can absorb more light energy because of the higher circumference-area ratio of smaller patches. 
  As depicted in Fig. \ref{fig4}(c.), the SIMU.normE result is quite different from the experimental result. The undulation (Profile 7 in Fig. \ref{fig4}(b.)) may be an important factor of the enhancement effect as mentioned above (see Fig. \ref{fig2}). Besides, a study showed that THz wave can heat probe and then change carrier density of substrate\cite{wiecha2021terahertz}. Based on the above facts, we propose the following assumptions:
\begin{enumerate}
      \item FIB etching implants high concentration $\rm{Ga^{+}}$ into Si-substrate.
      \item External electric field near the metals (tip apex, gold edges) is greatly enhanced. Electrons are heated and ejected into the silicon substrate. Heating of the substrate increase carrier density of the substrate.
      \item The increased carrier concentration makes the silicon substrate more conductive. However, its conductivity must be weaker than that of the gold film (if not, isolated and connected patches sharing the same signal strength), so the isolated patches (see Fig. \ref{fig4}(b.)) show a higher signal ratio (Eq. \ref{eq2}) than that in simulation (see Fig. \ref{fig4}(c.)). In addition, around the connected patches, carrier transport is easier, which results in larger signals.
      \item Besides, Joule heating and the carrier density increase will change substrate refractive index $n_s$. Subwavelength (aperture size $a$) metallic apertures resonant under the condition $\lambda/2\approx n_s a$ also cause signal enhancement\cite{kang2009substrate, adak2019nanoantenna}.
\end{enumerate}
  These explanations require more parameters for simulations and careful comparison with experiments.
  

\section{Methods}
\subsection{Sample Fabrication}
  Focused ion beam scanning electron microscopy (FIB-SEM, Helios G4 UX) is applied here to fabricate state-of-the-art nanometer 3D specimens and perform immediate high-resolution SEM imaging. Compared with electron-beam lithography and extreme-ultraviolet lithography, FIB-SEM can directly etch down the metal film and maintain a flat surface, while metal lift-off technology will introduce much unevenness and many defects on the metal surface, including at the sides \cite{seo2009terahertz}. In this recipe, first, a thin gold film (about $10~\rm{nm}$) is vaporized on a high-resistance silicon substrate (resistivity greater than $20~\rm{k\Omega\bullet cm}$, <111> orientation, thickness $512~\rm{\mu m}$). Second, FIB-SEM is used to fabricate the sample with in-situ SEM observation. The beam current is up to $1-7~\rm{pA}$ for accelerating voltages of $30~\rm{kV}$, under which the etching process only takes a few seconds to complete and a relatively uniform surface is maintained. A pattern generator is used to control the gallium ion action range, which may lead to severe damage in a narrow ditch during ion milling. The fabrication quality is checked by SEM and AFM. The resolution of the electron beam of the Helios G4 UX is better than $1.2~\rm{nm}$ and sufficient for the smallest structure in this experiment. The island growth and cross-section of the gold film are presented as insert images in Fig.\ref{fig1}(a.).
  
\subsection{Experimental Setup}
  A commercial TDS s-SNOM (Neaspec, GmbH, Germany) is applied in this experiment to scan the THz near-field optical signal and AFM topography. A tunable gas laser (SIFIR-50, Coherent Inc., U.S.A) operating between $0.2-2.5~\rm{THz}$ is used. The scattering signal is collected by THz bolometer (QMC Instruments Ltd., Cardiff, U.K.). A custom-designed probe with a length of up to $70~\mu m$ and a tip apex radius of up to $40~\rm{nm}$ is used to enhance the near-field effects. In this experiment, AFM is working in tapping mode, in which the tip will periodically contact the sample surface. The AFM tip scans an area of $(256\times256)$ pixels with a tip frequency $46.09~\rm{kHz}$ and a tapping amplitude of $175~\rm{nm}$.
  
\subsection{Simulation}
  Commercial FEM software (COMSOL Multiphysics, RF module) is used to numerically solve the field distribution of the real probe shape and sample structure. To reduce the calculation cost of the simulation, the metal probe shaft is hollowed out, impedance boundary or perfect electrical conductor (PEC) boundary conditions are used for metals, background electric field purely in Z-direction, and the results are calculated in the frequency domain. The simulation domain is a sphere (filled with air $\varepsilon'_{air}=1$) covered by a perfectly matched layer. High-resistance silicon (real part of the refractive index $n'_{air}=3.48$) is used only under the exposed substrate. The probe is a cone (shaft length of $70~ \rm{\mu m}$) with a sphere (tip apex radius of $40~ \rm{nm}$) at the top to avoid discontinuities. In the full wave simulation, a $2~ \rm{THz}$ p-polarized background field is applied to the $50~\rm{nm}$ tip-sample junction. All pixels of each sample are scanned, and the electric field distribution near the tip-sample junction is recorded. Internal self-grid division is used to mesh grids. The data are postprocessed with MATLAB into grayscale images.
 
  \subsubsection{Sample-1}
  For the noncontact scanning in simulation, a $7~\rm{\mu m}\times 7~\rm{\mu m}$ square area is adopted with a step length of $0.5~\rm{\mu m}$. The calculation for each pixel costs minutes on a common desktop computer requiring $40$ hours in total (desktop:128G memory, i7-10700 CPU) to draw a simulation image. The simulate-image of patches is performed separately to ensure sufficient pixels and less computation time.
  \subsubsection{Sample-2}
  For the noncontact scanning in simulation, a $2.2~\rm{\mu m}\times 2.2~\rm{\mu m}$ square area is adopted with a step length of $0.1~\rm{\mu m}$. The calculation for each pixel costs minutes on a common desktop computer requiring $37$ hours in total to draw a simulation image.

\section{Conclusions}
  THz nanoscopy of subwavelength structures are studied though experiments and simulations. THz-TDS SNOM provides near-field optical signals, which are related to the electric field distribution as  FEM simulation illustrated. Different from Drude model results, optical properties of sub-wavelength metal are related to confined electron transportation. Electron moving plays a vital role in tip-sample interaction and influences far-field detected signals. $1.4\times 1.4~\rm{{\mu m}^2}$ gold patches (Sample-1) and a more elaborate Sample-2 were examined. In THz-TDS s-SNOM nanoscopy, gold films are perfect conductors and simply determined by their connectivity and shapes. Comparing simulated images with experimental results, we are further understanding the electrical properties of tip-sample junction. The signal strength of microscale patches is well determined by the free electron density, and proportional to the patches' size. This give us an ideal reference to determine the carrier density in subwavelength structure. The $\rm{Ga^{+}}$ implanted Si-substrate have a higher signal than isolated patches, because of the larger area substrates have. However, this phenomenon is more complicated when we conduct experiments on nano-scale samples (comparable to tip apex radius). First of all, the silicon substrate is abnormally enhanced, even with higher signals than the peripheral gold film. Second, the signals from different patches are not solely related to their size. We believe that these effects are mainly caused by the $\rm{Ga^{+}}$ implantation. But other effects must participate in. We are looking for techniques to verify our conjecture and we believe the answer can help us further understand the THz nano-material interactions.

\begin{acknowledgement}
This work was supported by National Natural Science Foundation of China (Grant Nos. 61927813 and 61975225), Science and Technology Commission of Shanghai Municipality (Grant Nos. 21DZ1101102).
\end{acknowledgement}


\bibliography{achemso-demo}

\providecommand{\latin}[1]{#1}
\makeatletter
\providecommand{\doi}
  {\begingroup\let\do\@makeother\dospecials
  \catcode`\{=1 \catcode`\}=2 \doi@aux}
\providecommand{\doi@aux}[1]{\endgroup\texttt{#1}}
\makeatother
\providecommand*\mcitethebibliography{\thebibliography}
\csname @ifundefined\endcsname{endmcitethebibliography}
  {\let\endmcitethebibliography\endthebibliography}{}
\begin{mcitethebibliography}{42}
\providecommand*\natexlab[1]{#1}
\providecommand*\mciteSetBstSublistMode[1]{}
\providecommand*\mciteSetBstMaxWidthForm[2]{}
\providecommand*\mciteBstWouldAddEndPuncttrue
  {\def\EndOfBibitem{\unskip.}}
\providecommand*\mciteBstWouldAddEndPunctfalse
  {\let\EndOfBibitem\relax}
\providecommand*\mciteSetBstMidEndSepPunct[3]{}
\providecommand*\mciteSetBstSublistLabelBeginEnd[3]{}
\providecommand*\EndOfBibitem{}
\mciteSetBstSublistMode{f}
\mciteSetBstMaxWidthForm{subitem}{(\alph{mcitesubitemcount})}
\mciteSetBstSublistLabelBeginEnd
  {\mcitemaxwidthsubitemform\space}
  {\relax}
  {\relax}

\bibitem[Palik(1998)]{palik1998handbook}
Palik,~E.~D. \emph{Handbook of optical constants of solids}; Academic press,
  1998; Vol.~3\relax
\mciteBstWouldAddEndPuncttrue
\mciteSetBstMidEndSepPunct{\mcitedefaultmidpunct}
{\mcitedefaultendpunct}{\mcitedefaultseppunct}\relax
\EndOfBibitem
\bibitem[Auston and Nuss(1988)Auston, and Nuss]{auston1988electrooptical}
Auston,~D.~H.; Nuss,~M.~C. Electrooptical generation and detection of
  femtosecond electrical transients. \emph{IEEE Journal of quantum electronics}
  \textbf{1988}, \emph{24}, 184--197\relax
\mciteBstWouldAddEndPuncttrue
\mciteSetBstMidEndSepPunct{\mcitedefaultmidpunct}
{\mcitedefaultendpunct}{\mcitedefaultseppunct}\relax
\EndOfBibitem
\bibitem[Smith \latin{et~al.}(1988)Smith, Auston, and
  Nuss]{smith1988subpicosecond}
Smith,~P.~R.; Auston,~D.~H.; Nuss,~M.~C. Subpicosecond photoconducting dipole
  antennas. \emph{IEEE Journal of Quantum Electronics} \textbf{1988},
  \emph{24}, 255--260\relax
\mciteBstWouldAddEndPuncttrue
\mciteSetBstMidEndSepPunct{\mcitedefaultmidpunct}
{\mcitedefaultendpunct}{\mcitedefaultseppunct}\relax
\EndOfBibitem
\bibitem[Gallot \latin{et~al.}(1999)Gallot, Zhang, McGowan, Jeon, and
  Grischkowsky]{gallot1999measurements}
Gallot,~G.; Zhang,~J.; McGowan,~R.; Jeon,~T.-I.; Grischkowsky,~D. Measurements
  of the THz absorption and dispersion of ZnTe and their relevance to the
  electro-optic detection of THz radiation. \emph{Applied Physics Letters}
  \textbf{1999}, \emph{74}, 3450--3452\relax
\mciteBstWouldAddEndPuncttrue
\mciteSetBstMidEndSepPunct{\mcitedefaultmidpunct}
{\mcitedefaultendpunct}{\mcitedefaultseppunct}\relax
\EndOfBibitem
\bibitem[Cheon \latin{et~al.}(2016)Cheon, Yang, Lee, Kim, and
  Son]{cheon2016terahertz}
Cheon,~H.; Yang,~H.-j.; Lee,~S.-H.; Kim,~Y.~A.; Son,~J.-H. Terahertz molecular
  resonance of cancer DNA. \emph{Scientific Reports} \textbf{2016}, \emph{6},
  1--10\relax
\mciteBstWouldAddEndPuncttrue
\mciteSetBstMidEndSepPunct{\mcitedefaultmidpunct}
{\mcitedefaultendpunct}{\mcitedefaultseppunct}\relax
\EndOfBibitem
\bibitem[Hu and Nuss(1995)Hu, and Nuss]{hu1995imaging}
Hu,~B.~B.; Nuss,~M.~C. Imaging with terahertz waves. \emph{Optics Letters}
  \textbf{1995}, \emph{20}, 1716--1718\relax
\mciteBstWouldAddEndPuncttrue
\mciteSetBstMidEndSepPunct{\mcitedefaultmidpunct}
{\mcitedefaultendpunct}{\mcitedefaultseppunct}\relax
\EndOfBibitem
\bibitem[Wang \latin{et~al.}(2022)Wang, Xie, and Ying]{wang2022overview}
Wang,~Q.; Xie,~L.; Ying,~Y. Overview of imaging methods based on terahertz
  time-domain spectroscopy. \emph{Applied Spectroscopy Reviews} \textbf{2022},
  \emph{57}, 249--264\relax
\mciteBstWouldAddEndPuncttrue
\mciteSetBstMidEndSepPunct{\mcitedefaultmidpunct}
{\mcitedefaultendpunct}{\mcitedefaultseppunct}\relax
\EndOfBibitem
\bibitem[Qiu \latin{et~al.}(2022)Qiu, You, Tan, Wan, Wang, Liu, Chen, Liu, Tao,
  Fu, \latin{et~al.} others]{qiu2022terahertz}
Qiu,~F.; You,~G.; Tan,~Z.; Wan,~W.; Wang,~C.; Liu,~X.; Chen,~X.; Liu,~R.;
  Tao,~H.; Fu,~Z., \latin{et~al.}  A terahertz near-field nanoscopy revealing
  edge fringes with a fast and highly sensitive quantum-well photodetector.
  \emph{Iscience} \textbf{2022}, \emph{25}, 104637\relax
\mciteBstWouldAddEndPuncttrue
\mciteSetBstMidEndSepPunct{\mcitedefaultmidpunct}
{\mcitedefaultendpunct}{\mcitedefaultseppunct}\relax
\EndOfBibitem
\bibitem[Maissen \latin{et~al.}(2019)Maissen, Chen, Nikulina, Govyadinov, and
  Hillenbrand]{maissen2019probes}
Maissen,~C.; Chen,~S.; Nikulina,~E.; Govyadinov,~A.; Hillenbrand,~R. Probes for
  ultrasensitive THz nanoscopy. \emph{Acs Photonics} \textbf{2019}, \emph{6},
  1279--1288\relax
\mciteBstWouldAddEndPuncttrue
\mciteSetBstMidEndSepPunct{\mcitedefaultmidpunct}
{\mcitedefaultendpunct}{\mcitedefaultseppunct}\relax
\EndOfBibitem
\bibitem[Mastel \latin{et~al.}(2018)Mastel, Govyadinov, Maissen, Chuvilin,
  Berger, and Hillenbrand]{mastel2018understanding}
Mastel,~S.; Govyadinov,~A.~A.; Maissen,~C.; Chuvilin,~A.; Berger,~A.;
  Hillenbrand,~R. Understanding the image contrast of material boundaries in IR
  nanoscopy reaching 5 nm spatial resolution. \emph{ACS Photonics}
  \textbf{2018}, \emph{5}, 3372--3378\relax
\mciteBstWouldAddEndPuncttrue
\mciteSetBstMidEndSepPunct{\mcitedefaultmidpunct}
{\mcitedefaultendpunct}{\mcitedefaultseppunct}\relax
\EndOfBibitem
\bibitem[Walther \latin{et~al.}(2002)Walther, Plochocka, Fischer, Helm, and
  Uhd~Jepsen]{walther2002collective}
Walther,~M.; Plochocka,~P.; Fischer,~B.; Helm,~H.; Uhd~Jepsen,~P. Collective
  vibrational modes in biological molecules investigated by terahertz
  time-domain spectroscopy. \emph{Biopolymers: Original Research on
  Biomolecules} \textbf{2002}, \emph{67}, 310--313\relax
\mciteBstWouldAddEndPuncttrue
\mciteSetBstMidEndSepPunct{\mcitedefaultmidpunct}
{\mcitedefaultendpunct}{\mcitedefaultseppunct}\relax
\EndOfBibitem
\bibitem[Shen \latin{et~al.}(2021)Shen, Zhu, Zhang, Guo, Zhang, Ren, Chen, Li,
  and Zhao]{shen2021ultra}
Shen,~J.; Zhu,~Z.; Zhang,~Z.; Guo,~C.; Zhang,~J.; Ren,~G.; Chen,~L.; Li,~S.;
  Zhao,~H. Ultra-broadband terahertz fingerprint spectrum of melatonin with
  vibrational mode analysis. \emph{Spectrochimica Acta Part A: Molecular and
  Biomolecular Spectroscopy} \textbf{2021}, \emph{247}, 119141\relax
\mciteBstWouldAddEndPuncttrue
\mciteSetBstMidEndSepPunct{\mcitedefaultmidpunct}
{\mcitedefaultendpunct}{\mcitedefaultseppunct}\relax
\EndOfBibitem
\bibitem[Ordal \latin{et~al.}(1983)Ordal, Long, Bell, Bell, Bell, Alexander,
  and Ward]{ordal1983optical}
Ordal,~M.~A.; Long,~L.; Bell,~R.; Bell,~S.; Bell,~R.; Alexander,~R.; Ward,~C.
  Optical properties of the metals al, co, cu, au, fe, pb, ni, pd, pt, ag, ti,
  and w in the infrared and far infrared. \emph{Applied Optics} \textbf{1983},
  \emph{22}, 1099--1119\relax
\mciteBstWouldAddEndPuncttrue
\mciteSetBstMidEndSepPunct{\mcitedefaultmidpunct}
{\mcitedefaultendpunct}{\mcitedefaultseppunct}\relax
\EndOfBibitem
\bibitem[Benavides-Cruz \latin{et~al.}(2016)Benavides-Cruz,
  Calder{\'o}n-Ram{\'o}n, Gomez-Aguilar, Rodr{\'\i}guez-Achach,
  Cruz-Ordu{\~n}a, Laguna-Camacho, Morales-Mendoza, Enciso-Aguilar,
  P{\'e}rez-Meana, Escalante-Mart{\'\i}nez, \latin{et~al.}
  others]{benavides2016numerical}
Benavides-Cruz,~M.; Calder{\'o}n-Ram{\'o}n,~C.; Gomez-Aguilar,~J.;
  Rodr{\'\i}guez-Achach,~M.; Cruz-Ordu{\~n}a,~I.; Laguna-Camacho,~J.;
  Morales-Mendoza,~L.; Enciso-Aguilar,~M.; P{\'e}rez-Meana,~H.;
  Escalante-Mart{\'\i}nez,~J., \latin{et~al.}  Numerical simulation of metallic
  nanostructures interacting with electromagnetic fields using the
  Lorentz--Drude model and FDTD method. \emph{International Journal of Modern
  Physics C} \textbf{2016}, \emph{27}, 1650043\relax
\mciteBstWouldAddEndPuncttrue
\mciteSetBstMidEndSepPunct{\mcitedefaultmidpunct}
{\mcitedefaultendpunct}{\mcitedefaultseppunct}\relax
\EndOfBibitem
\bibitem[Jeon \latin{et~al.}(2004)Jeon, Kim, Kang, Maeng, Son, An, Lee, and
  Lee]{jeon2004optical}
Jeon,~T.-I.; Kim,~K.-J.; Kang,~C.; Maeng,~I.~H.; Son,~J.-H.; An,~K.~H.;
  Lee,~J.~Y.; Lee,~Y.~H. Optical and electrical properties of preferentially
  anisotropic single-walled carbon-nanotube films in terahertz region.
  \emph{Journal of Applied Physics} \textbf{2004}, \emph{95}, 5736--5740\relax
\mciteBstWouldAddEndPuncttrue
\mciteSetBstMidEndSepPunct{\mcitedefaultmidpunct}
{\mcitedefaultendpunct}{\mcitedefaultseppunct}\relax
\EndOfBibitem
\bibitem[Pandey \latin{et~al.}(2016)Pandey, Gupta, Chanana, and
  Nahata]{pandey2016non}
Pandey,~S.; Gupta,~B.; Chanana,~A.; Nahata,~A. Non-Drude like behaviour of
  metals in the terahertz spectral range. \emph{Advances in Physics: X}
  \textbf{2016}, \emph{1}, 176--193\relax
\mciteBstWouldAddEndPuncttrue
\mciteSetBstMidEndSepPunct{\mcitedefaultmidpunct}
{\mcitedefaultendpunct}{\mcitedefaultseppunct}\relax
\EndOfBibitem
\bibitem[Park \latin{et~al.}(2009)Park, Choi, Ahn, Rotermund, Sohn, Kang,
  Jeong, and Kim]{park2009terahertz}
Park,~D.; Choi,~S.; Ahn,~Y.; Rotermund,~F.; Sohn,~I.; Kang,~C.; Jeong,~M.;
  Kim,~D. Terahertz near-field enhancement in narrow rectangular apertures on
  metal film. \emph{Optics Express} \textbf{2009}, \emph{17},
  12493--12501\relax
\mciteBstWouldAddEndPuncttrue
\mciteSetBstMidEndSepPunct{\mcitedefaultmidpunct}
{\mcitedefaultendpunct}{\mcitedefaultseppunct}\relax
\EndOfBibitem
\bibitem[Chen \latin{et~al.}(2013)Chen, Park, Pelton, Piao, Lindquist, Im, Kim,
  Ahn, Ahn, Park, \latin{et~al.} others]{chen2013atomic}
Chen,~X.; Park,~H.-R.; Pelton,~M.; Piao,~X.; Lindquist,~N.~C.; Im,~H.;
  Kim,~Y.~J.; Ahn,~J.~S.; Ahn,~K.~J.; Park,~N., \latin{et~al.}  Atomic layer
  lithography of wafer-scale nanogap arrays for extreme confinement of
  electromagnetic waves. \emph{Nature Communications} \textbf{2013}, \emph{4},
  1--7\relax
\mciteBstWouldAddEndPuncttrue
\mciteSetBstMidEndSepPunct{\mcitedefaultmidpunct}
{\mcitedefaultendpunct}{\mcitedefaultseppunct}\relax
\EndOfBibitem
\bibitem[Moon \latin{et~al.}(2015)Moon, Park, Kim, Do, Lee, Lee, Kang, and
  Han]{moon2015subsurface}
Moon,~K.; Park,~H.; Kim,~J.; Do,~Y.; Lee,~S.; Lee,~G.; Kang,~H.; Han,~H.
  Subsurface nanoimaging by broadband terahertz pulse near-field microscopy.
  \emph{Nano Letters} \textbf{2015}, \emph{15}, 549--552\relax
\mciteBstWouldAddEndPuncttrue
\mciteSetBstMidEndSepPunct{\mcitedefaultmidpunct}
{\mcitedefaultendpunct}{\mcitedefaultseppunct}\relax
\EndOfBibitem
\bibitem[Chen \latin{et~al.}(2020)Chen, Liu, Guo, Chen, Hu, Nikulina, Ye, Yao,
  Bechtel, Martin, \latin{et~al.} others]{chen2020thz}
Chen,~X.; Liu,~X.; Guo,~X.; Chen,~S.; Hu,~H.; Nikulina,~E.; Ye,~X.; Yao,~Z.;
  Bechtel,~H.~A.; Martin,~M.~C., \latin{et~al.}  THz near-field imaging of
  extreme subwavelength metal structures. \emph{ACS Photonics} \textbf{2020},
  \emph{7}, 687--694\relax
\mciteBstWouldAddEndPuncttrue
\mciteSetBstMidEndSepPunct{\mcitedefaultmidpunct}
{\mcitedefaultendpunct}{\mcitedefaultseppunct}\relax
\EndOfBibitem
\bibitem[Cocker \latin{et~al.}(2021)Cocker, Jelic, Hillenbrand, and
  Hegmann]{cocker2021nanoscale}
Cocker,~T.; Jelic,~V.; Hillenbrand,~R.; Hegmann,~F. Nanoscale terahertz
  scanning probe microscopy. \emph{Nature Photonics} \textbf{2021}, \emph{15},
  558--569\relax
\mciteBstWouldAddEndPuncttrue
\mciteSetBstMidEndSepPunct{\mcitedefaultmidpunct}
{\mcitedefaultendpunct}{\mcitedefaultseppunct}\relax
\EndOfBibitem
\bibitem[Huber \latin{et~al.}(2008)Huber, Keilmann, Wittborn, Aizpurua, and
  Hillenbrand]{2008Terahertz}
Huber,~A.~J.; Keilmann,~F.; Wittborn,~J.; Aizpurua,~J.; Hillenbrand,~R.
  Terahertz near-field nanoscopy of mobile carriers in single semiconductor
  nanodevices. \emph{Nano Letters} \textbf{2008}, \emph{8}, 3766\relax
\mciteBstWouldAddEndPuncttrue
\mciteSetBstMidEndSepPunct{\mcitedefaultmidpunct}
{\mcitedefaultendpunct}{\mcitedefaultseppunct}\relax
\EndOfBibitem
\bibitem[Hillenbrand \latin{et~al.}(2002)Hillenbrand, Taubner, and
  Keilmann]{hillenbrand2002phonon}
Hillenbrand,~R.; Taubner,~T.; Keilmann,~F. Phonon-enhanced light--matter
  interaction at the nanometre scale. \emph{Nature} \textbf{2002}, \emph{418},
  159--162\relax
\mciteBstWouldAddEndPuncttrue
\mciteSetBstMidEndSepPunct{\mcitedefaultmidpunct}
{\mcitedefaultendpunct}{\mcitedefaultseppunct}\relax
\EndOfBibitem
\bibitem[Cvitkovic \latin{et~al.}(2007)Cvitkovic, Ocelic, and
  Hillenbrand]{cvitkovic2007analytical}
Cvitkovic,~A.; Ocelic,~N.; Hillenbrand,~R. Analytical model for quantitative
  prediction of material contrasts in scattering-type near-field optical
  microscopy. \emph{Optics Express} \textbf{2007}, \emph{15}, 8550--8565\relax
\mciteBstWouldAddEndPuncttrue
\mciteSetBstMidEndSepPunct{\mcitedefaultmidpunct}
{\mcitedefaultendpunct}{\mcitedefaultseppunct}\relax
\EndOfBibitem
\bibitem[Knoll and Keilmann(2000)Knoll, and Keilmann]{knoll2000enhanced}
Knoll,~B.; Keilmann,~F. Enhanced dielectric contrast in scattering-type
  scanning near-field optical microscopy. \emph{Optics Communications}
  \textbf{2000}, \emph{182}, 321--328\relax
\mciteBstWouldAddEndPuncttrue
\mciteSetBstMidEndSepPunct{\mcitedefaultmidpunct}
{\mcitedefaultendpunct}{\mcitedefaultseppunct}\relax
\EndOfBibitem
\bibitem[Chen \latin{et~al.}(2019)Chen, Hu, Mescall, You, Basov, Dai, and
  Liu]{chen2019modern}
Chen,~X.; Hu,~D.; Mescall,~R.; You,~G.; Basov,~D.; Dai,~Q.; Liu,~M. Modern
  scattering-type scanning near-field optical microscopy for advanced material
  research. \emph{Advanced Materials} \textbf{2019}, \emph{31}, 1804774\relax
\mciteBstWouldAddEndPuncttrue
\mciteSetBstMidEndSepPunct{\mcitedefaultmidpunct}
{\mcitedefaultendpunct}{\mcitedefaultseppunct}\relax
\EndOfBibitem
\bibitem[Pizzuto \latin{et~al.}(2021)Pizzuto, Chen, Hu, Dai, Liu, and
  Mittleman]{pizzuto2021anomalous}
Pizzuto,~A.; Chen,~X.; Hu,~H.; Dai,~Q.; Liu,~M.; Mittleman,~D.~M. Anomalous
  contrast in broadband THz near-field imaging of gold microstructures.
  \emph{Optics Express} \textbf{2021}, \emph{29}, 15190--15198\relax
\mciteBstWouldAddEndPuncttrue
\mciteSetBstMidEndSepPunct{\mcitedefaultmidpunct}
{\mcitedefaultendpunct}{\mcitedefaultseppunct}\relax
\EndOfBibitem
\bibitem[Zhang \latin{et~al.}(2021)Zhang, Hu, Zhang, Wang, Zhang, Xu, Zhao, Wu,
  Zhong, Liu, \latin{et~al.} others]{zhang2021direct}
Zhang,~Z.; Hu,~M.; Zhang,~X.; Wang,~Y.; Zhang,~T.; Xu,~X.; Zhao,~T.; Wu,~Z.;
  Zhong,~R.; Liu,~D., \latin{et~al.}  Direct observation of tip-gap
  interactions in THz scattering-type scanning near-field optical microscopy.
  \emph{Applied Physics Express} \textbf{2021}, \emph{14}, 102004\relax
\mciteBstWouldAddEndPuncttrue
\mciteSetBstMidEndSepPunct{\mcitedefaultmidpunct}
{\mcitedefaultendpunct}{\mcitedefaultseppunct}\relax
\EndOfBibitem
\bibitem[Huth \latin{et~al.}(2013)Huth, Chuvilin, Schnell, Amenabar,
  Krutokhvostov, Lopatin, and Hillenbrand]{huth2013resonant}
Huth,~F.; Chuvilin,~A.; Schnell,~M.; Amenabar,~I.; Krutokhvostov,~R.;
  Lopatin,~S.; Hillenbrand,~R. Resonant antenna probes for tip-enhanced
  infrared near-field microscopy. \emph{Nano Letters} \textbf{2013}, \emph{13},
  1065--1072\relax
\mciteBstWouldAddEndPuncttrue
\mciteSetBstMidEndSepPunct{\mcitedefaultmidpunct}
{\mcitedefaultendpunct}{\mcitedefaultseppunct}\relax
\EndOfBibitem
\bibitem[Chen \latin{et~al.}(2017)Chen, Lo, Zheng, Hu, Dai, and
  Liu]{chen2017rigorous}
Chen,~X.; Lo,~C. F.~B.; Zheng,~W.; Hu,~H.; Dai,~Q.; Liu,~M. Rigorous numerical
  modeling of scattering-type scanning near-field optical microscopy and
  spectroscopy. \emph{Applied Physics Letters} \textbf{2017}, \emph{111},
  223110\relax
\mciteBstWouldAddEndPuncttrue
\mciteSetBstMidEndSepPunct{\mcitedefaultmidpunct}
{\mcitedefaultendpunct}{\mcitedefaultseppunct}\relax
\EndOfBibitem
\bibitem[McArdle \latin{et~al.}(2020)McArdle, Lahneman, Biswas, Keilmann, and
  Qazilbash]{mcardle2020near}
McArdle,~P.; Lahneman,~D.; Biswas,~A.; Keilmann,~F.; Qazilbash,~M. Near-field
  infrared nanospectroscopy of surface phonon-polariton resonances.
  \emph{Physical Review Research} \textbf{2020}, \emph{2}, 023272\relax
\mciteBstWouldAddEndPuncttrue
\mciteSetBstMidEndSepPunct{\mcitedefaultmidpunct}
{\mcitedefaultendpunct}{\mcitedefaultseppunct}\relax
\EndOfBibitem
\bibitem[Chen \latin{et~al.}(2021)Chen, Yao, Xu, McLeod, Gilbert~Corder, Zhao,
  Tsuneto, Bechtel, Martin, Carr, \latin{et~al.} others]{chen2021hybrid}
Chen,~X.; Yao,~Z.; Xu,~S.; McLeod,~A.~S.; Gilbert~Corder,~S.~N.; Zhao,~Y.;
  Tsuneto,~M.; Bechtel,~H.~A.; Martin,~M.~C.; Carr,~G.~L., \latin{et~al.}
  Hybrid machine learning for scanning near-field optical spectroscopy.
  \emph{ACS Photonics} \textbf{2021}, \emph{8}, 2987--2996\relax
\mciteBstWouldAddEndPuncttrue
\mciteSetBstMidEndSepPunct{\mcitedefaultmidpunct}
{\mcitedefaultendpunct}{\mcitedefaultseppunct}\relax
\EndOfBibitem
\bibitem[McLeod \latin{et~al.}(2014)McLeod, Kelly, Goldflam, Gainsforth,
  Westphal, Dominguez, Thiemens, Fogler, and Basov]{mcleod2014model}
McLeod,~A.~S.; Kelly,~P.; Goldflam,~M.; Gainsforth,~Z.; Westphal,~A.~J.;
  Dominguez,~G.; Thiemens,~M.~H.; Fogler,~M.~M.; Basov,~D. Model for
  quantitative tip-enhanced spectroscopy and the extraction of
  nanoscale-resolved optical constants. \emph{Physical Review B} \textbf{2014},
  \emph{90}, 085136\relax
\mciteBstWouldAddEndPuncttrue
\mciteSetBstMidEndSepPunct{\mcitedefaultmidpunct}
{\mcitedefaultendpunct}{\mcitedefaultseppunct}\relax
\EndOfBibitem
\bibitem[Mooshammer \latin{et~al.}(2020)Mooshammer, Huber, Sandner, Plankl,
  Zizlsperger, and Huber]{mooshammer2020quantifying}
Mooshammer,~F.; Huber,~M.~A.; Sandner,~F.; Plankl,~M.; Zizlsperger,~M.;
  Huber,~R. Quantifying nanoscale electromagnetic fields in near-field
  microscopy by Fourier demodulation analysis. \emph{Acs Photonics}
  \textbf{2020}, \emph{7}, 344--351\relax
\mciteBstWouldAddEndPuncttrue
\mciteSetBstMidEndSepPunct{\mcitedefaultmidpunct}
{\mcitedefaultendpunct}{\mcitedefaultseppunct}\relax
\EndOfBibitem
\bibitem[Feres \latin{et~al.}(2021)Feres, Mayer, Wehmeier, Maia, Viana,
  Malachias, Bechtel, Klopf, Eng, Kehr, \latin{et~al.} others]{feres2021sub}
Feres,~F.~H.; Mayer,~R.~A.; Wehmeier,~L.; Maia,~F.~C.; Viana,~E.;
  Malachias,~A.; Bechtel,~H.~A.; Klopf,~J.~M.; Eng,~L.~M.; Kehr,~S.~C.,
  \latin{et~al.}  Sub-diffractional cavity modes of terahertz hyperbolic phonon
  polaritons in tin oxide. \emph{Nature Communications} \textbf{2021},
  \emph{12}, 1--9\relax
\mciteBstWouldAddEndPuncttrue
\mciteSetBstMidEndSepPunct{\mcitedefaultmidpunct}
{\mcitedefaultendpunct}{\mcitedefaultseppunct}\relax
\EndOfBibitem
\bibitem[Zhang \latin{et~al.}(2021)Zhang, Zhang, Wang, and
  Chen]{zhang2021detection}
Zhang,~W.; Zhang,~K.; Wang,~W.; Chen,~Y. Detection of ion implantation in
  focused ion beam processing by scattering-type scanning near-field optical
  microscopy. \emph{Optics Letters} \textbf{2021}, \emph{46}, 649--652\relax
\mciteBstWouldAddEndPuncttrue
\mciteSetBstMidEndSepPunct{\mcitedefaultmidpunct}
{\mcitedefaultendpunct}{\mcitedefaultseppunct}\relax
\EndOfBibitem
\bibitem[Graziosi \latin{et~al.}(2018)Graziosi, Mi, Kiss, and
  Quack]{graziosi2018single}
Graziosi,~T.; Mi,~S.; Kiss,~M.; Quack,~N. Single crystal diamond micro-disk
  resonators by focused ion beam milling. \emph{Apl Photonics} \textbf{2018},
  \emph{3}, 126101\relax
\mciteBstWouldAddEndPuncttrue
\mciteSetBstMidEndSepPunct{\mcitedefaultmidpunct}
{\mcitedefaultendpunct}{\mcitedefaultseppunct}\relax
\EndOfBibitem
\bibitem[Wiecha \latin{et~al.}(2021)Wiecha, Kapoor, and
  Roskos]{wiecha2021terahertz}
Wiecha,~M.~M.; Kapoor,~R.; Roskos,~H.~G. Terahertz scattering-type near-field
  microscopy quantitatively determines the conductivity and charge carrier
  density of optically doped and impurity-doped silicon. \emph{APL Photonics}
  \textbf{2021}, \emph{6}, 126108\relax
\mciteBstWouldAddEndPuncttrue
\mciteSetBstMidEndSepPunct{\mcitedefaultmidpunct}
{\mcitedefaultendpunct}{\mcitedefaultseppunct}\relax
\EndOfBibitem
\bibitem[Kang \latin{et~al.}(2009)Kang, Choe, Kim, and Park]{kang2009substrate}
Kang,~J.; Choe,~J.-H.; Kim,~D.; Park,~Q.-H. Substrate effect on aperture
  resonances in a thin metal film. \emph{Optics Express} \textbf{2009},
  \emph{17}, 15652--15658\relax
\mciteBstWouldAddEndPuncttrue
\mciteSetBstMidEndSepPunct{\mcitedefaultmidpunct}
{\mcitedefaultendpunct}{\mcitedefaultseppunct}\relax
\EndOfBibitem
\bibitem[Adak and Tripathi(2019)Adak, and Tripathi]{adak2019nanoantenna}
Adak,~S.; Tripathi,~L.~N. Nanoantenna enhanced terahertz interaction of
  biomolecules. \emph{Analyst} \textbf{2019}, \emph{144}, 6172--6192\relax
\mciteBstWouldAddEndPuncttrue
\mciteSetBstMidEndSepPunct{\mcitedefaultmidpunct}
{\mcitedefaultendpunct}{\mcitedefaultseppunct}\relax
\EndOfBibitem
\bibitem[Seo \latin{et~al.}(2009)Seo, Park, Koo, Park, Kang, Suwal, Choi,
  Planken, Park, Park, \latin{et~al.} others]{seo2009terahertz}
Seo,~M.; Park,~H.; Koo,~S.; Park,~D.; Kang,~J.; Suwal,~O.; Choi,~S.;
  Planken,~P.; Park,~G.; Park,~N., \latin{et~al.}  Terahertz field enhancement
  by a metallic nano slit operating beyond the skin-depth limit. \emph{Nature
  Photonics} \textbf{2009}, \emph{3}, 152--156\relax
\mciteBstWouldAddEndPuncttrue
\mciteSetBstMidEndSepPunct{\mcitedefaultmidpunct}
{\mcitedefaultendpunct}{\mcitedefaultseppunct}\relax
\EndOfBibitem
\end{mcitethebibliography}

\end{document}